\documentclass{sigkddExp}
\usepackage{graphicx}
\usepackage{amsmath}
\usepackage[caption=false]{subfig}
\graphicspath{ {images/} }
\usepackage[labelfont=bf,textfont=bf]{caption}
\usepackage{makecell}
\usepackage{flushend}
\usepackage{cite}
\usepackage{booktabs}
\usepackage{multirow}
\usepackage{siunitx}
\usepackage{colortbl,dcolumn}

\definecolor{green2}{rgb}{.23,.59,.17}
\definecolor{green4}{rgb}{.17,.73,.09}
\definecolor{green1}{rgb}{0.169,0.51,0.316}
\definecolor{green3}{rgb}{0.075,0.6236,0.301}

\begin{document}
\title{Learning to Rank Broad and Narrow Queries in E-Commerce}

\numberofauthors{3}

\author{Siddhartha Devapujula\\Myntra\\siddhartha.devapujula@myntra.com \and     Sagar Arora\\Myntra\\sagar.arora@myntra.com \and Sumit Borar\\Google\\sumitborar@gmail.com\thanks{Work done while at Myntra}}

\date{25 October 2018}
\maketitle
\begin{abstract}

Search is a prominent channel for discovering products on an e-commerce platform. Ranking products retrieved from search becomes crucial to address customer's need and optimize for business metrics. While learning to Rank (LETOR) models have been extensively studied and have demonstrated efficacy in the context of web search; it is a relatively new research area to be explored in the e-commerce.

In this paper, we present a framework for building LETOR model for an e-commerce platform. We analyze user queries and propose a mechanism to segment queries between broad and narrow based on user's intent. We discuss different types of features - query, product and query-product and discuss challenges in using them. We show that sparsity in product features can be tackled through a denoising auto-encoder while skip-gram based word embeddings help solve the query-product sparsity issues. We also present various target metrics that can be employed for evaluating search results and compare their robustness. 

Further, we build and compare performances of both pointwise and pairwise LETOR models on fashion category data set. We also build and compare distinct models for broad and narrow queries, analyze feature importance across these and show that these specialized models perform better than a combined model in the fashion world.

\end{abstract}



\section{Introduction}

Users on an e-commerce platform typically discover products through search, browsing categories or marketing campaigns. On our platform, search functionality is key to product discovery as each of these channels translates to a search query in the back-end. Search ranking is a critical aspect of our business. Hence any  improvement in the product search ranking, enhances user's experience on the platform and results in better business metrics. Although Learning to Rank (LETOR) has demonstrated the efficacy in the context of web search; it is a relatively new research area to be explored in the e-commerce. One major difference between the web search and e-commerce search lies in the interplay of relevance and popularity features. For example, e-commerce queries like ``tshirts" can yield a recall set of thousands of products which make it indispensable to incorporate popularity features. Also, typical human relevance judgments would not suffice as it would in the context of web search \cite{kumar2018did,karmaker2017application}. Section \ref{related-work} describes more about the work in this area.\par \noindent

Furthermore, in categories like fashion or home goods products lack a distinctive name and a unique identification is not possible. Users typically describe their intent in a search query using product attributes, ex: for purchasing a \textit{Tshirt} user could search for \textit{\{tshirt\}} or \textit{\{\textless brand \textgreater  t-shirts\}} or \textit{\{\textless color \textgreater t-shirts\}}. Each of these queries would typically result in thousands of relevant products from the catalogue as opposed to queries in hard goods categories with well defined products like ``iphone X 64 gb" which would result in few relevant products.
We refer such queries like \textit{\{tshirt\}} as \textit{unnamed queries}. Further exacerbating this situation there exists an infinite virtual shelf space in e-commerce and limited real-estate on user's devices. Therefore, it becomes vital to optimally rank products to enable efficient product discovery. The paper discusses the approach and challenges involved in applying LETOR methods to address search rankings for \textit{unnamed queries} in e-commerce domain.



\begin{figure}[h]
\centering
\includegraphics[height=1.6in, width=2.2in]{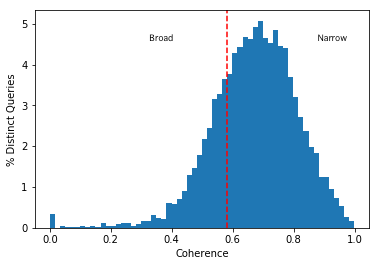}
\caption{Query Distribution basis Coherency Scores}
\label{figure:hist}
\end{figure}

For such queries on our platform we observe a more skewed distribution of traffic across queries, a 90-10 distribution than typical 80-20 distribution where 10\% of queries lead to 90\% of the traffic i.e. number of downstream sessions as a result of query. Also, we observed that traffic and recall set size are strongly correlated. So, the few queries that lead to high traffic would also have a very large recall set. We also inspected the average click depth \footnote{The mean search engine ranking position for all the clicked products} of our queries, which showed significantly different user behavior. This provides a motivation to segment queries into high recall set (aka broad) and low recall set (aka narrow) for the ranking purposes. While the narrow queries have a clear intent and result in low average click depth, the broad queries are the ones which are browse intensive and have large average click depth. Examples of queries which constitute as broad queries are sports shoes, wallets for men, dresses, forever 21 dresses and of narrow queries are longline jacket for men, high rise ripped jeans for women, Blackberrys Charcoal Trousers, deal jeans long sleeves tops, open toe ballerinas etc. We describe our approach and analysis on broad and narrow queries in Section \ref{broad-narrow}.

\begin{figure}[h]
\centering
\includegraphics[height=1.6in, width=2.2in]{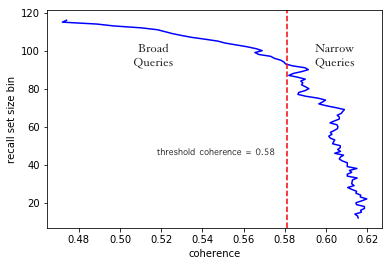}
\caption{Coherency Score vs Recall Set Size}
\label{figure:coh-recall}
\end{figure}

Section 4 describes the system architecture and designs and Section 5 discusses the feature engineering and target variables. Finally, Section 6 describes our modeling approach, results and analysis. Our paper makes the following contributions:
\begin{enumerate}
\item We provide an approach to segment our queries into broad and narrow basis how coherent the downstream sessions are. We show that segmenting queries and training different models for each can be a better approach than training single model across the board
    \item Apart from using typical query features, product features and query-product features, we propose a denoising autoencoder based architecture to reduce sparsity of product features and skip gram based word embeddings for query-product features. 
    \item We demonstrate the impact of various combinations of different types of features on the model's performance. Further, we study the behaviour of various target variables - CTR, Add to cart ratio, conversion and Revenue Per Impression.
    \item We highlight the differences between broad and narrow queries in terms of modelling approach, feature importance etc. We also show how our model significantly improves NDCG compared to the baseline model built upon style popularity.
\end{enumerate}

\section{Related Work} \label{related-work}

LETOR methods have demonstrated their success in web search \cite{chapelle2011yahoo, azzopardi2016advances}. Various LETOR models like RankNet, LamdaMart, AdaRank and RankBoost have been compared on Web Search data \cite{li2014learning,  liu2009learning,  burges2010ranknet,  chapelle2011yahoo}.  Moreover search has been extensively studied in the e-commerce primarily from the retrieval perspective \cite{agrawal2009diversifying, li2011towards, long2012enhancing, van2016learning, yu2014latent}. 

 Karmaker et al.  \cite{karmaker2017application}  attempted to apply LETOR in the e-commerce \footnote{Walmart dataset} setting. They showed how LamdaMart is able to optimize the combination of popularity and relevance based features; and perform the best even in E-commerce ranking. The work also suggested that it might be helpful to segment queries and train separate models for each segment; however it ended up training one model across board.

Researchers have often segregated queries in web search domain too.  Kumaripaba Athukorala et al's presented an interesting work on Broad and Narrow queries in web search \cite{athukorala2014narrow}. Also, Cheng Luo et al. \cite{luo2014query} defines 3 types of queries: 
\textit{Ambiguous}: a query that has more than one meaning 
\textit{Broad}: a query that covers a variety of subtopics, and a user might look for one of the subtopics by issuing another query
\textit{Clear}: a query that has a specific meaning and covers a narrow topic. Both claim that supporting exploratory search is a very challenging problem,
because of the dynamic nature of the exercise and intent of the users. In e-commerce domain, we tackle this problem by building a classification model as described in Section \ref{broad-narrow}. 

\section{Broad and Narrow Queries} \label{broad-narrow}

We randomly sampled 100k queries and labelled them as Broad/Narrow. We have divided the queries into train and test in the ratio of 70:30. Later we have trained a SVM classifier with radial-basis kernel on the train queries. We have considered multiple sets of features - Word2Vec of query, result set size and identified query attributes.

\subsection{Coherency score} \label{coherency-socre}

For every query based on the recall set, we define coherency score which is used as a feature for the classification model. We create an embedding of each product on the platform using Matrix Factorization atop matrix of all user-product clicks on our platform. For a query $q$, and recall set $R = {P_1,P_2,...P_n}$ , we define centroid $R_c$ and coherency score $c_{qr}$ as 

\begin{equation}
    R_c = \frac{\sum P_i}{n}
\end{equation}

\begin{equation}
    c_{qr} = \underset{i}{\mathrm{median}} \{<p_i.R_c>\}
\end{equation}

It is the median of inner product of each product with the centroid of all products. Typically, queries with large recall set would contribute to majority of traffic on our platform and would also have low coherency score.

\subsection{Word2Vec Embeddings} \label{word2vec-embeddings}

We use query clickstream logs for training the word embeddings. It is a dataset of all the queries and subsequent downstream sessions as a result of query (specifically the products which were clicked, added to cart and finally purchased). As in \cite{arora2016decoding}, we consider each session i.e. sequence of products $P_1,P_2, ....P_n$ as the document, and the attributes of each product (eg brand-nike, color-red, neck-polo) as the words. Directly using each product as words can induce sparsity because products are separated in the time space. This document is further appended with query words and top-10 synonyms of each word in the query (to augment our vocabulary) from wordnet \cite{miller1995wordnet}. Skip-gram model learns embedding of each query token, external synonyms from wordnet and all the catalogue attributes. The product embedding is computed as the centroid of attribute embeddings. And the query embedding is centroid of all unigrams in query.

Formally, let $q \ \epsilon \  Q$ be a query from the query corpus. $W$ is the set of all unigrams in each query with $S$ as the set of top-10 synonyms of each unigram in query. Further, let $P_1,P_2,...P_n$ be the products clicked in a particular session as a result of the query and $A_i$ be the set of attributes of $P_i$. Then a document $d$ is constructed as:
$d = W \cup S \cup \{A_i\} (1 \leq i \leq n)$ 
We construct such documents from all query-sessions in our clickstream and train skip-gram model on the same to learn embedding for each attribute ($v_{a_i}$) / query unigram ($v_{w_i}$) / query synonym ($v_{s_i}$). 

\begin{equation}
    v_{p_{i}} = \frac{\sum_{a_i \epsilon A_i } {v_{a_i}}}{n} 
\end{equation}
\begin{equation}
    v_{q_{i}} = \frac{\sum_{w_i \epsilon W}   {v_{w_i}}}{|W|}   
\end{equation}

\subsection{Query Attributes} \label{query-attributes}
Some of the query features include basic attributes like query length, number of words in query, etc. We have fragmented the query and also included features such as the presence of the attributes. Ex: whether the query contains an article type or brand or color.

\subsection{Target Variable} \label{target-classifier}

In order to train the model, we have assigned the labels by a formula based on heuristics which is primarily based on the coherency score. Typically, queries with large recall set would contribute to majority of traffic on our platform and would also have low coherency score.

\begin{table}[hb]
\centering
\begin{tabular}{|l|l|l|l|l|l|}
\hline
\textbf{Metric} & \textbf{Broad} & \textbf{Narrow} \\ \hline
Percent Queries             & 13 \%            & 87 \%                         \\ \hline
Percent Traffic (Sessions)           & 73 \%            & 27 \%                         \\ \hline
Distinct Brands          & 15.2            & 4.26                         \\ \hline
Distinct Article Types          & 2.7            & 1.14                         \\ \hline

\end{tabular}
\caption{Query Segment Statistics}
\label{table:stats}
\end{table}

The figure \ref{figure:hist} shows  the distribution of coherency score (number of distinct queries corresponding to each score) and figure \ref{figure:coh-recall} shows the recall set bin corresponding to each coherency score. We had divided all the recall set sizes into 120 uniform-sized bins for this experiment. We threshold at coherency score of 0.58, basis the elbow shown in figure \ref{figure:coh-recall}. This corresponds to bin number 92; which further corresponds to a recall set size of 1910-2300. Clearly, it can be attributed to \textit{unnamed queries}. So, a query with coherency score $\leq$ 0.58 is referred to as broad query while  query with coherency score $>$ 0.58 is referred to as narrow query. The table \ref{table:stats} shows various statistics regarding both the segments. It is evident that very few broad queries lead to majority of traffic. Also, broad queries often have more number of distinct brands and article types the narrow queries. In order to avoid the outliers, we have considered only queries with coherency scores $\leq$ 0.3 as broad queries and $>$ 0.7 as narrow queries from the train set, rest all queries are filtered out.

\subsection{Performance} \label{performance}

We used accuracy as an indication of model's performance. We generated the predictions on the test queries which achieved an overall accuracy of 0.812. Apart from the test data we further manually evaluated a sample of  10k queries by the experts and the accuracy on those queries was found out to be 0.89. The SVM classifier outperformed the primary heuristics baseline model which had an accuracy of 0.83.

\section{System Design} \label{system-design}

\begin{figure}[ht]
\includegraphics[height=1.8in, width=3in]{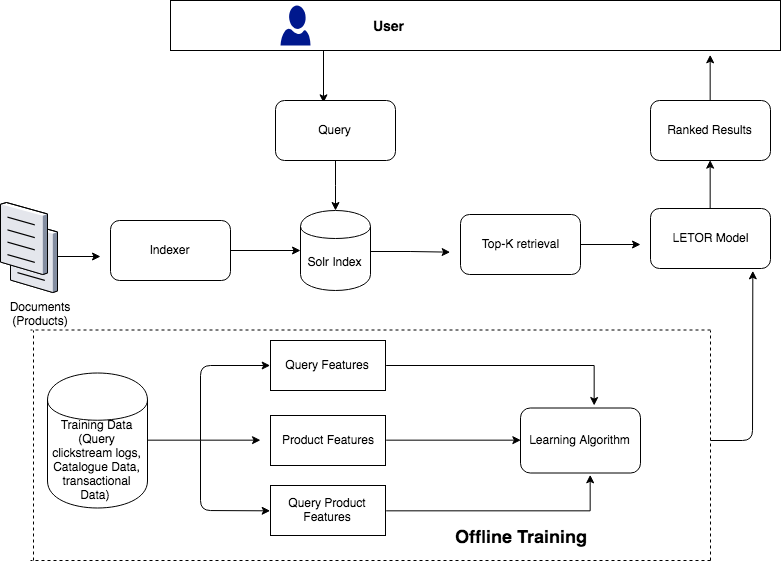}
\caption{Architecture}
\label{figure:arch}
\end{figure}

This section discusses the architecture (figure \ref{figure:arch}) involved in retrieval and ranking of search products on our platform. We use SOLR as the underlying search engine with over 3M fashion products indexed. 

When a user issues a \textit{query}, the retrieval layer renders a set of top-K (typically 1000) products based on BM-25 score. Traditional BM25 based approaches are quite effective for retrieval; however in broad queries like ``tshirts", it becomes extremely significant to include business metrics like CTR and conversion to optimize the ranking. Not just the demand, it becomes critical for our platform to include the available inventory as an input to the ranking algorithm. To incorporate that we have a basic model which assigns a global score to each product as a weighted combination of various features related to product's performance in past 2 weeks (CTR, revenue, quantity sold etc.) and inventory. \footnote{The primary focus of this paper would be LETOR, instead} Final search recall score is computed as weighted sum of BM25 score and the product performance score. The top-K results are further re-ranked using the trained LETOR model. It is worth mentioning that we have separate models for broad and narrow queries. The LETOR model is invoked accordingly.
The LETOR model is trained using:
\begin{enumerate}
    \item Catalogue Data : Structured information regarding each product's  physical features like  brand, color, mrp etc. 
    \item Transactional Data: Product's output business metrics like daywise revenue, CTR etc.
    \item Query-Clickstream logs : Logs each query and information regarding query's downstream sessions like products seen (impressions), clicked, added to cart, wishlisted, liked, purchased etc.
\end{enumerate}

\section{Model}

Learning to rank is a popular approach that provides a principled way to  optimize ranking of search results given various features. This section focuses on various features and  target variables  we used for training our LETOR models. From modelling perspective, we tried 2 pointwise models - Random Forests and Gradient Boosting Model and 2 pairwise models - RankNet and LamdaMart. We built three different types of models - one specific for broad queries, one specific for narrow queries and lastly a combined model across the board. 
 
\subsection{Feature Engineering} \label{feat-eng}

We use three different types of features:

\begin{enumerate}
    \item \textbf{Query Features}: These are features specific to query like total length of query, number of words, is brand (eg Nike, Tommy Hilfiger) present in query, is article type (eg Dresses, Shoes) present in query, the identified article type, brand etc.

    \item \textbf{Product Features}: These are features specific to the products (documents) They can either be popularity related or physical features. The popularity features include features involving past performance of the product's brand or article type (hereafter referred to as entity) like revenue in 15 days, quantity sold in 15 days etc. It is worth mentioning that we do not directly include the performance metrics of each product directly, as suggested in \cite{karmaker2017application}. We observed that this leads to  to overfitting and poor generalisation on unseen products. 
    
    The physical features include feature types which are uniform across article types (eg brand, color, material/fabric etc) and those which are specific to article types. For instance tshirts would have attributes like neck (round/polo) while shoes could have attributes like outsole type (marking/non-marking), cleated (yes/no) etc. The total number of distinct attributes (key-value pairs) on our platform is over 10k. These features are quite sparse in nature and models trained using them don't perform well. We trained a separate denoising autoencoder to reduce the dimensionality of product's physical features. 
    
    The architecture is shown in figure \ref{figure:autoencoder} We consider products from each master category (for eg all tshirts, casual shirts, jackets, sweaters etc. would fall in topwear) and provide one-hot encoded representation of its attributes as the input to the network. A uniform noise from 0 to 1 is added for better generalisation. Post that we have 3 layered encoder and decoder as shown in figure \ref{figure:autoencoder}. The final learnt embeddings consists of 32 dimensions. This greatly captures and embeds the most informative features in the original feature representation as we discuss later in the Results Section where we compare this with normal autoencoder (without the noise layer). The autoencoder based approach would greatly help to reduce sparsity in features and further assist LETOR models to optimise ranking by learning weights (parameters) for different features.
    
    \begin{figure}
\includegraphics[height=1.8in, width=3in]{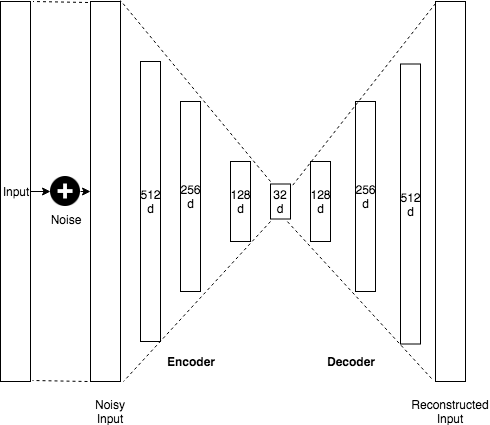}
\caption{Denoising Autoencoder Architecture}
\label{figure:autoencoder}
\end{figure}

\item \textbf{Query Product Features} The query product features are features which involve both query and product - eg ctr of a tshirt product when the query is ``batman printed tshirt". The query product features can again be of 2 different types: popularity based and relevance based. The popularity features include past performance of product's entity as a result of query. Again, we do not directly include query-product features on product level (eg revenue of product from given query) to avoid overfitting.

The relevance related query product features encompasses features describing the match or similarity between given query and product. BM 25 is a traditional example for the same. To better capture the similarity between query and product, we attempt to embed all our queries and products in a common ``query relevance" space using skip-gram based embeddings as described in Section \ref{word2vec-embeddings}. And the query product relevance score is computed as the inner product of query embedding and the product embedding.

    \end{enumerate}

    \subsection{Targets} \label{target}
    
   As demonstrated in \cite{kumar2018did}, while effective for web search, crowdsourcing is not a very effective way to obtain relevance judgments for E-com search. This can be attributed to the large recall set in e-commerce and how difficult it is to describe a fashion product - making it indispensable to incorporate business metrics as the relevance.
   
   For a query $q$, and product $p$, let $I_{qp}$ represent the total number of impressions, $C_{qp}$ be the total numbers clicks, $B_{qp}$ be the total number of add to bag / cart, $Q_{qp}$ be the total products purchased and $R_{qp}$ be the total revenue of product as result of the query. We use the following target variables for our analysis:
   
   \begin{enumerate}
       \item \textbf{Click-Through Rate} It is the probability of clicking on the listpage (the page as a result of search query). Better the ranking, higher would be the CTR
        \begin{equation}
            CTR_{qp} = \frac{C_{qp}}{I_{qp}} 
        \end{equation} 
        
        \item \textbf{Add to cart Rate} It is the probability of adding a product to the cart post the click. It is the perceived utility of click page.
        \begin{equation}
            ATCR_{qp} = \frac{B_{qp}}{C_{qp}} 
        \end{equation} 
        
        \item \textbf{Conversion} It is the probability of purchasing a product from listpage. This can be considered as the overall satisfaction of the user.
        \begin{equation}
            Conv_{qp} = \frac{B_{qp}}{I_{qp}} 
        \end{equation}
        
        \item \textbf{Revenue per Impression} It refers to the overall business value (revenue) from each impression as a result of the query.
        \begin{equation}
            RPI_{qp} = \frac{R_{qp}}{I_{qp}} 
        \end{equation}

   \end{enumerate}

    \section{Results}
    
    \subsection{Dataset}

  We randomly sampled 100k queries. We have classified them into broad and narrow using on the model described in Section 3. We resulted 13k broad queries and 87k narrow queries. We sampled queries in 80-20 proportion in stratified manner to collate train (broad and narrow) and test (broad and narrow) queries. 
  

\subsubsection{Train and Test Data}

We use the clickstream logs to  retrieve the list of products which had impressions when the query was fired (for every train query). Every $<query-product>$  pair is represented as a row. For modeling purposes, we pruned the rows with very few or very large number of impressions. We used $\log(ctr)$ as the target variable (computed for each query-product  pair) for all our experiments. Initially we built a model over all the queries and later experimented with two different models on broad and narrow queries separately.
  
For testing purposes, top 500 products are retrieved for the test queries based on the existing search ranking (as discussed in Section \ref{system-design}) and the features are extracted for all the query-product pairs. We predict $\log(ctr)$ for the query-product and  evaluate the performance using NDCG metric as discussed in next section.

      \subsubsection{Evaluation Metric}

NDCG is the metric used most often in the web search. We used the same in our context; and report the average NDCG over all queries. In our case, the ideal ranking of products is inferred from the product's past performance given the query context. We considered the true realized ctr (or other metrics like atcr, conversion, rpi) of query-product to compute the ideal ranking of products. As in \cite{karmaker2017application}, the ground truth relevance ratings based on the target variable were computed for all products by normalizing and discretizing to a $b$ point integer scale ${1,2,..b}$; where b is the bin size. The relevance score for ground truth ranking in case of ctr can be computed as:

\begin{equation}
    rel_{ctr_{qp}} = ceil(\frac{b.ctr_{qp}}{max_{p|q} ctr_{qp}})
\end{equation}

For computing NDCG@K, we take the top k (we used $K=48$) products for the query and compute relevance scores for each query-product as above. 

Let $i_p$ be the predicted rank position for each product and $i_i$ be the ideal rank position. Now for each query $q$, compute 

\begin{equation}
    DCG = \sum_p \frac{2^{rel_{qp} - 1}}{log_2(i_p+1)} 
\end{equation}

\begin{equation}
    IDCG = \sum_p \frac{2^{rel_{qp} - 1}}{log_2(i_i+1)} 
\end{equation}

\begin{equation}
    NDCG = \frac{DCG}{IDCG}
\end{equation}

    \subsection{Cross Target Learning}

In e-commerce, choosing one target variable for LETOR is not easy; especially considering the large correlation between the target variables. We intend to optimize for multiple business metrics (discussed in Section \ref{target}) simultaneously. We  evaluated the performance of LambdaMART models trained using one training target on test datasets based on other target. Also, we have typically transformed all our target variables using logarithm because of an approximate gaussian distribution. The results are tabulated in Table \ref{table:target-study}. We observe that all train metrics perform well when tested on the same metric but experience a drop when tested on others (diagonal entries in Table \ref{table:target-study}). Conversion metric performs the best in terms of robustness, since it performs well across the board. This can be attributed to how conversion captures the entire funnel of user's behaviour on platform (from visit to click, cart, order etc.) However, conversion faces sparsity issues because very few products (amongst the entire catalogue) end up being converted. Hence, we recommend to use CTR to build models initially and transition to conversion once we have enough data. All our experimental results are based on CTR metric.
\begin{table}[ht]
\begin{tabular}{|l|l|l|l|l|}
\hline
\multicolumn{1}{|c|}{\multirow{2}{*}{\textbf{Train Target}}} & \multicolumn{4}{c|}{\textbf{Test Target}}                    \\ 
\multicolumn{1}{|c|}{}                                       & \textbf{\newline CTR} & \textbf{ATCR} & \textbf{Conv} & \textbf{RPI} \\ \hline
\textbf{log(ctr)}                                            &  0.79&  0.71& 0.69 & 0.71          \\ \hline
\textbf{log(atcr)}                                            & 0.75& 0.82& 0.71 & 0.73          \\ \hline
\textbf{log(conversion)}                                            & 0.71& 0.79& 0.78 & 0.78          \\ \hline
\textbf{log(rpi)}                                            & 0.68 & 0.75 & 0.73 &	0.76          \\ \hline
\end{tabular}
\caption{Cross Target Learning}
\label{table:target-study}
\end{table}



\subsection{Feature Study}
    
    
    
    
\subsubsection{Reducing Sparsity in Query Product Features}

We took all queries over 15 days for training the skip-gram model with window size 10. We randomly sampled 10M documents (each document is query and corresponding session).  The table \ref{table:w2v} shows few similar attributes to some queries along with cosine similarity. We use these skip-gram embeddings to compute query-product relevance score as described in Section \ref{feat-eng}

\begin{figure}[h]
\includegraphics[height=1.8in, width=3in]{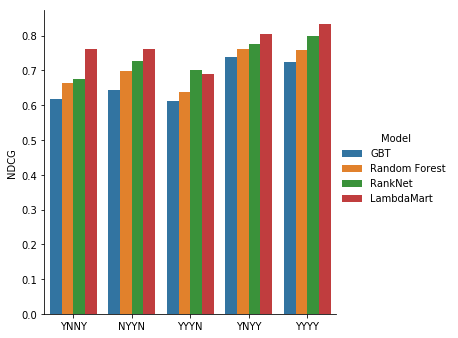}
\caption{Comparison of different models. All xlabels are explained in Table \ref{table:different_combinations}}
\label{figure:modelcompare}
\end{figure}
     
 \subsubsection{Feature Combinations and Models}
For all our experiments and analysis, we divided features into 4 different types - query features, query-product features (includes both word2vec based relevance and popularity features), product popularity features and product autoencoder based (physical) features. We experimented with 5 different combinations of features as shown in Table \ref{table:different_combinations}

\begin{table}[b]
\centering
\begin{tabular}{|p{1.5cm}|p{5cm}|}
\hline
\textbf{Query} & \textbf{Similar Attributes with cosine similarity}  \\ \hline
nike & dri-fit(0.72), adidas(0.64), puma(0.58), sportwear(0.55)    \\ \hline
baniyan (hindi word for vest) & vests white(0.57), sando (0.513), cotton vest (0.51), innerwear(0.504)    \\ \hline
swimwear & swimsuit(0.915), swimdress(0.718), tankini(0.701), bikini(0.606)    \\ \hline
party dress & maxis (0.752), short dress (0.674), gown(0.663), black dress(0.65),     \\ \hline
\end{tabular}
\caption{Attributes similar to query}
\label{table:w2v}
\end{table}
     
\begin{table}[t]
\centering
\begin{tabular}{|p{2cm}|p{1.2cm}|p{1.2cm}|p{1.2cm}|p{1.2cm}|}
\hline
\textbf{Name} & \textbf{Query Features} & \textbf{Query Product Features} & \textbf{Product Popularity Features} & \textbf{Product Physical Features} \\ \hline
Only Relevance features & Y & N & N & Y   \\ \hline
Only Popularity features & N & Y & Y & N   \\ \hline
No Query Product features & Y & N & Y & Y    \\ \hline
No Autoencoder features  & Y & Y & Y & N    \\ \hline
All features & Y & Y & Y & Y    \\ \hline
\end{tabular}
\caption{Different Feature Combinations}
\label{table:different_combinations}
\end{table}

We will constantly use the codes (eg YNYY) mentioned in Table \ref{table:different_combinations} hereafter.
        
Firstly, we compare different models (2 pointwise - Random Forest, GBM and 2 pairwise - RankNet and LamdaMart). The figure \ref{figure:modelcompare} shows the NDCG of different models across different feature combinations. 
        
 \begin{figure}[hb]
\includegraphics[height=1.8in, width=3in]{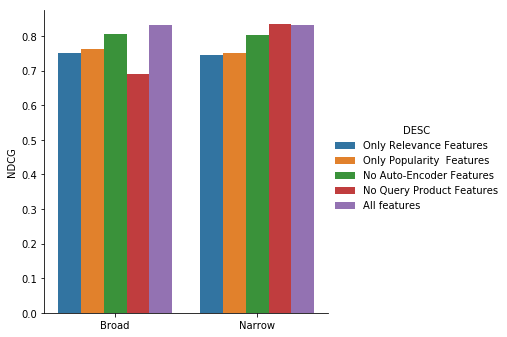}
\caption{Broad Vs Narrow}
\label{figure:broadvsnarrow}
\end{figure}       

Clearly, LamdaMart performs the best across different feature combinations (except YYYN where RankNet performs marginally better). Considering LamdaMart as the model, we can also study the performance of different feature combinations in figure \ref{figure:modelcompare}. In terms of NDCG performance, $YYYN < YNNY < NYYN < YNYY < YYYY$ Furthermore, introducing autoencoder features helped a lot; as is evident from gain in $YYYN \ to \   YYYY$. Also, popularity features (both product and query-product) helps to improve the performance. We will be studying the importance of each type of feature across broad and narrow queries later in this section.

Next, we attempted to see if training separate models for broad and narrow queries help to improve performance than training just a single model across the board. For this experiment, we used LamdaMart model and feature combination as YYYY.
The table \ref{table:combined} shows the test NDCG values. Clearly, training separate models for broad and narrow queries boost the overall performance of broad and narrow queries. The improvement in narrow queries (12 \%) is much higher than that in broad queries (2.4 \%). 

\begin{table}
\centering
\begin{tabular}{|p{3cm}|p{1.5cm}|p{1.5cm}|}
\hline
\textbf{NDCG} & \textbf{Broad Queries} & \textbf{Narrow Queries} \\ \hline
Single Model & 0.81 & 0.74   \\ \hline
Model on Broad Queries & 0.83 & 0.72   \\ \hline
Model on Narrow Queries & 0.79 & 0.83  \\ \hline
\end{tabular}
\caption{Combined vs Individual Model}
\label{table:combined}
\end{table}

We also compared our final model (built separately on broad and narrow queries) to the baseline model (discussed in Section \ref{system-design}). The table \ref{table:baseline} compares our LETOR model with the baseline

\begin{table}
\centering
\begin{tabular}{|p{3cm}|p{1.5cm}|p{1.5cm}|}
\hline
\textbf{Query Type} & \textbf{Baseline} & \textbf{LETOR} \\ \hline
Broad & 0.75 & 0.83  \\ \hline
Narrow & 0.69  & 0.83   \\ \hline
\end{tabular}
\caption{LETOR vs Popularity Based Model}
\label{table:baseline}
\end{table}

We observe that improvement in narrow queries (20.28 \%)  is significantly higher than broad queries (10.67 \%). This can be attributed the role of popularity features in broad features. While popularity plays the major role in broad queries, relevance has a major role to play in narrow queries where the user's intent is quite clear. 





Finally, we study the importance of different types of features for broad and narrow queries. Figure \ref{figure:broadvsnarrow} demonstrates how inclusion and exclusion of features affect the performance. Query style popularity features have a important role in the model built over broad queries compared to the one built over narrow queries. Exclusion of this set of features showed a sharp decline in the NDCG. This shows we are able to learn the ranking from the product physical features and query features for narrow queries more accurately. This could be because of the more information in the query and the specific intent of the user.

\section{Conclusions}

We presented a framework for building LETOR models for an e-commerce platform - specifically for the \textit{unnamed queries}. We proposed a notion of coherency score and used it to segment queries into broad and narrow. We discussed the challenges involved in feature representation (query, product and query-product) and target metrics (ctr,atcr,conversion,rpi) in e-commerce. We proposed a denoising auto-encoder based architecture to tackle sparsity in product features and skip-gram based word embeddings for query-product sparsity issues. We compared the performances of both pointwise and pairwise LETOR models on fashion category data set.  We discussed how segmenting queries into broad and narrow (basis ) help in LETOR models for e-commerce. While popularity has a major role to play in broad queries, it is not that significant in narrow queries. We reported NDCG@48 for all the queries, it might be more relevant decide K based on the nature of the query (broad or narrow).

 In future, we intend to remove inherit bias (position, presentation bias etc) from the targets while training. When a new brand is launched, typically the brand based popularity will have zero values and the products of the brand will be penalized. Such cold start problems are to be resolved. We plan to build multiple models by strategically varying features and targets and later combine the predictions by an ensemble model.



\balancecolumns

\clearpage

%
\bibliographystyle{IEEEtran}
\bibliography{sigproc}  
%
%
\end{document}